# Octave Spanning Visible to SWIR Integrated Coil-Stabilized Brillouin Lasers

MEITING SONG[1], NITESH CHAUHAN[2, 3], MARK W. HARRINGTON[1], NICK MONTIFIORE[1], KAIKAI LIU[1], ANDREW S. HUNTER[1], CHRIS CARON[4], ANDREI ISICHENKO[1], ROBERT J. NIFFENEGGER[4], DANIEL J. BLUMENTHAL[1,*]

[1]Department of Electrical and Computer Engineering, University of California Santa Barbara, Santa Barbara, CA 93106, USA

[2]Time and Frequency Division, National Institute of Standards and Technology, Boulder, CO 80305, USA

[3]Department of Physics, University of Colorado, Boulder, CO 80309, USA

[4]Department of Electrical and Computer Engineering, University of Massachusetts Amherst, Amherst, MA 01003, USA

*Corresponding author, danb@ucsb.edu

**Abstract:** Narrow linewidth stabilized lasers are central to precision applications that operate across the visible to short-wave infrared wavelengths, including optical clocks, quantum sensing and computing, ultra-low noise microwave generation, and fiber sensing. Today, these spectrally pure sources are realized using multiple external cavity tabletop lasers locked to bulk-optic free-space reference cavities. Integration of this technology will enable portable precision applications with improved reliability and robustness. Here, we report wavelength-flexible design and operation, over more than an octave span, of an integrated coil-resonator-stabilized Brillouin laser architecture. Leveraging a versatile two-stage noise reduction approach, we achieve low linewidths and high stability with chip-scale laser designs based on the ultra-low-loss, CMOS-compatible silicon nitride platform. We report operation at 674 and 698 nm for applications to strontium neutral and trapped-ion clocks, quantum sensing and computing, and at 1550 nm for applications to fiber sensing and ultra-low phase noise microwave generation. Over this range we demonstrate frequency noise reduction from 1 Hz to 10 MHz resulting in 1.0 Hz -17 Hz fundamental and 181 Hz - 630 Hz integral linewidths and an Allan deviation of 6.5 x $10^{-13}$ at 1 ms for 674 nm, 6.0 x $10^{-13}$ at 15 ms for 698 nm, and 2.6 x$10^{-13}$ at 15 ms for 1550 nm. This represents the lowest achieved linewidths and highest stability for integrated stabilized Brillouin lasers over an order of magnitude improvement in operating wavelength range. These results unlock the potential of integrated, ultra-low-phase-noise stabilized lasers for precision applications and further integration in systems-on-chip solutions.

## Introduction

Stabilized narrow-linewidth lasers, operating across a wide range of visible (VIS) to short-wave infrared (SWIR) wavelengths, are essential tools for applications such as neutral atom and trapped-ion based quantum sensing and computing[1–3], ultra-low phase noise microwave and mmWave generation[4–7], and fiber sensing[8,9]. These applications can benefit from integration of today's lab-scale stabilized lasers, leading to a reduction in size, cost, and power consumption and improvement in reliability and stability. Integrated solutions that can replace these lasers must meet the required low phase noise and carrier stability while at the same time support laser emission and reference cavity operation that by design can span over an octave wavelength range from the VIS to SWIR.

Today, precision systems are constructed using SWIR fiber lasers or external cavity diode lasers (ECDL) that are Pound Drever-Hall (PDH) locked to bulk-optic ultra-low expansion (ULE) reference cavities[10–14]. The table-top lasers reduce the mid- to high-frequency offset noise while the reference cavity is used to mitigate close-to-carrier noise and improve carrier stability. To accommodate VIS light applications these solutions incorporate bulk-optic or waveguide second harmonic generation (SHG)[15]. Progress has been made on miniaturization of reference cavities using micro-optic designs[16–18], yet such solutions are not amenable to photonic integration, do not support agile tuning for stabilization to atomic references, and have large GHz-scale free spectral range (FSR) that do not support locking of fine resolution wavelength tuning.

Integrated approaches can employ dual stage laser noise reduction using a first stage integrated laser with nonlinear suppression of high frequency offset noise followed by a second cavity stage that reduces low- to mid-frequency offset noise[19]. Integrated lasers that employ non-linear feedback mechanisms include stimulated Brillouin scattering (SBS) and self-injection locking, and have been demonstrated in the VIS[20–24] to SWIR[25–28]. Integrated stabilization cavities such as coil or spiral resonators have the large optical mode volume needed to reduce the low- to mid-frequency noise down to the thermorefractive noise (TRN) limit[29,30]. The silicon nitride platform offers low loss from the VIS to SWIR[25,31–33], and can be used to realize Brillouin lasers and coil resonator stabilization cavities that can operate across octave wavelengths, with application to a wide range of precision applications. However, Brillouin laser stabilization using a common CMOS compatible platform for both laser and reference cavity with design flexibility to operate across a more than octave spanning range, has yet to be demonstrated.

Here, we report a class of integrated stabilized Brillouin lasers, capable of operation by design from the VIS to SWIR, fabricated in the CMOS compatible low-loss $Si_3N_4$ integration platform. Over an octave span, we demonstrate orders of magnitude frequency noise reduction from 1 Hz to 10 MHz, resulting in sub-20 Hz fundamental linewidth (FLW) and sub-kHz $1/\pi$ integral linewidth (ILW) emission with specific operation at 674 nm for $Sr^+$ ion applications[15,34], 698 nm for Sr neutral atom applications[2], and 1550 nm for fiber sensing and ultra-low noise microwave applications[4,8]. The concept and potential applications are illustrated in Fig. 1. Frequency noise and linewidth reduction are achieved in two stages, first using SBS in $Si_3N_4$ resonators with the appropriate wavelength pump laser to lower high frequency offset noise and then PDH locking the SBS laser to a meter-scale $Si_3N_4$ coil-resonator reference operating at the same wavelength for low- to mid-frequency offset noise reduction. Using this approach, we reduce the pump laser FLW to Hz-level and the ILW to sub-10 kHz level, across over an octave range. The meter-scale coil resonators are enabled by ultra-low losses and ultra-high quality factors (Q)[4]. We measure 0.63 dB/m and $Q_i = 94 \times 10^6$ at $\lambda = 674$ nm, 0.53 dB/m and $Q_i = 110 \times 10^6$ at $\lambda = 698$ nm, and 0.64 dB/m and $Q_i = 41 \times 10^6$ at $\lambda = 1550$ nm (Supplementary Information S1). The combination of dilute waveguide modes and coil resonator length provides a large optical mode volume with low TRN floor[30] which enables a 10x integral linewidth reduction over the TRN limit of the SBS laser alone. These low losses also enable direct emission SBS lasing with thresholds of 7 mW, 5 mW, and 14.6 mW at 674 nm, 698 nm, and 1550 nm respectively. In the visible, SBS is used to generate a 14 Hz FLW at 674 nm and a FLW of 7 Hz at 698 nm, while in the SWIR the SBS FLW is 1.0 Hz at 1550 nm. The meter-scale coil resonator provides an additional order of magnitude ILW reduction for each Brillouin laser, measuring 322 Hz, 630 Hz, and 181 Hz ILW at 674 nm, 698 nm, and 1550 nm respectively. The coil-resonators also provide short term laser stability with the Allan Deviation (ADEV) measured at $6.5 \times 10^{-13}$ at 1 ms for 674 nm, $6.0 \times 10^{-13}$ at 15 ms for 698 nm, and $2.6 \times 10^{-13}$ at 15 ms for 1550 nm. The ability to reduce the frequency noise and improve stability across the VIS to SWIR, by design, holds promise for integrated precision lasers across a wide range of applications to unlock the potential to move precision experiments out of the laboratory and into the field and portable applications.

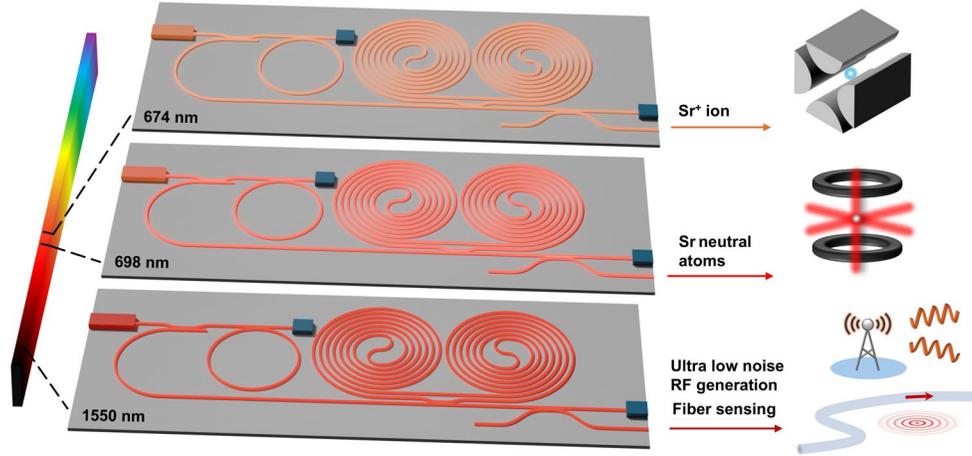

**Fig. 1** Illustration of integrated photonic coil-stabilized Brillouin lasers used for generation ultra-low phase noise and high stability at 674 nm for Sr+ ions, 698 nm for Sr neutral atoms, and 1550 nm for long-distance fiber sensing and ultra-low phase noise RF and microwave generation.

## Results

We demonstrate operation across over an octave from 674 nm to 1550 nm of integrated SBS lasers stabilized to integrated coil-resonators using PDH locking. We achieve orders of magnitude frequency noise reduction across the whole frequency range 1 Hz to 10 MHz with close to TRN floor limited operation. The stabilized laser (Fig. 2a) operates by first locking the pump laser to the integrated SBS resonator, enabling the continuous generation of first Stokes (S1) emission. The Stokes output is optically amplified and frequency tuned using an acoustic-optic modulator (AOM) for locking to the coil resonator. In the future, the AOM stage can be removed by direct tuning of the SBS laser[35]. The S1 output is locked to the integrated coil resonator using a PDH loop and the AOM for frequency noise reduction. The two stages of linewidth narrowing are illustrated in Fig. 2b and 2c, where Brillouin lasing reduces the high frequency offset noise and reduces the fundamental (Lorentzian) linewidth of the laser and the second stage lowers the low- to mid- frequency offset noise and reduces the integral linewidth of the laser. The fundamental linewidth is constrained by a combination of SBS laser intracavity power, resonator quality factor (Q), and required operation below the onset of second-order Stokes lasing. The integral linewidth narrowing is limited by the TRN floor of the coil resonator and the gain and bandwidth of the PDH lock.

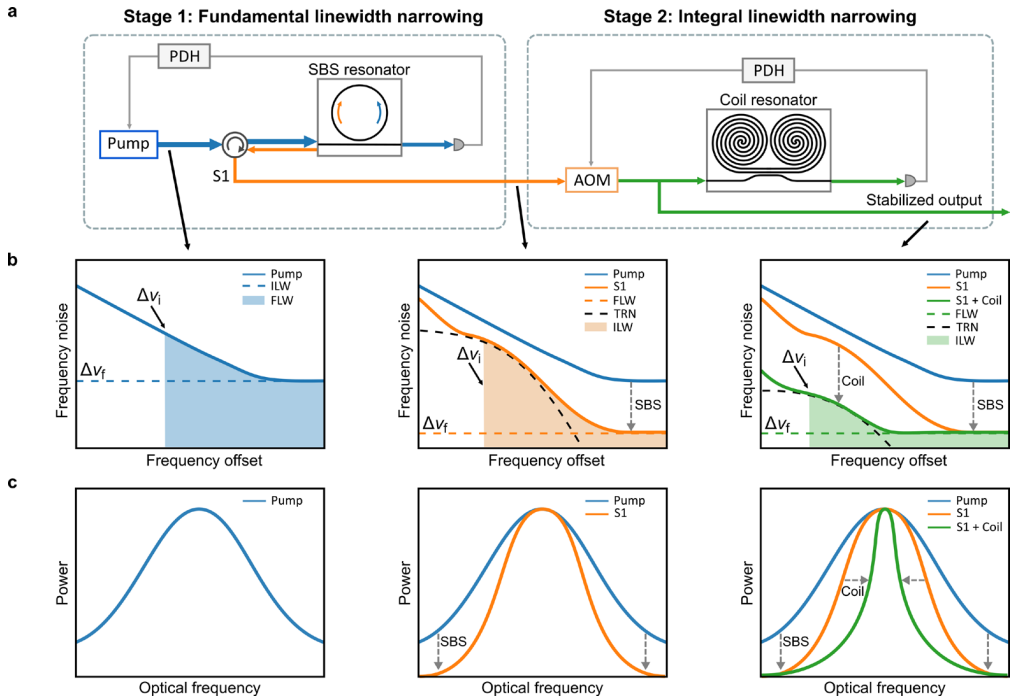

**Fig. 2 Coil-stabilized SBS laser and concept of two-stage linewidth narrowing. a.** Schematic of coil-stabilized SBS experiment. Blue represents the pump laser, orange represents the S1 emission, and green represents the coil-stabilized SBS laser. **b.** Illustration of high frequency noise reduction of pump laser noise using SBS and low- to mid-frequency noise reduction using coil-stabilization. The TRN limit (black dashed curve), fundamental linewidth (FLW, green dashed line), and integral linewidth (ILW, green shaded area) are shown. **c.** Illustration of linewidth narrowing of the pump laser, S1 emission, and coil-stabilized SBS laser.

The coil-resonators and SBS lasers are realized using a weakly confining thin-core, low-loss, high-Q $Si_3N_4$ design[25,31,36]. Here, we demonstrate that the same 40 nm thick nitride waveguide core can be used for both SBS lasers and coil resonators in the visible with mask-only changes in the waveguide width. At 1550 nm, the SBS laser is 40 nm thick for lower loss and threshold, while the coil resonator is 80 nm thick for increased compactness. Benefits of this waveguide configuration and fabrications details can be found in Ref. 36. The bend radii are kept above the critical bend radius and the absorption loss is lowered using high-temperature annealing. The TM mode is chosen to reduce sidewall scattering loss[31]. The SBS resonator waveguide width is 2.3 μm for 674 nm, 2.5 μm for 698 nm, and 11 μm for 1550 nm. The coil resonator waveguide widths are 3 μm for 674 nm and 698 nm, and 6 μm for 1550 nm. The losses are measured using resonator Q factor (Supplementary Information S1) and a calibrated Mach-Zehnder Interferometer (MZI) method (see Methods).

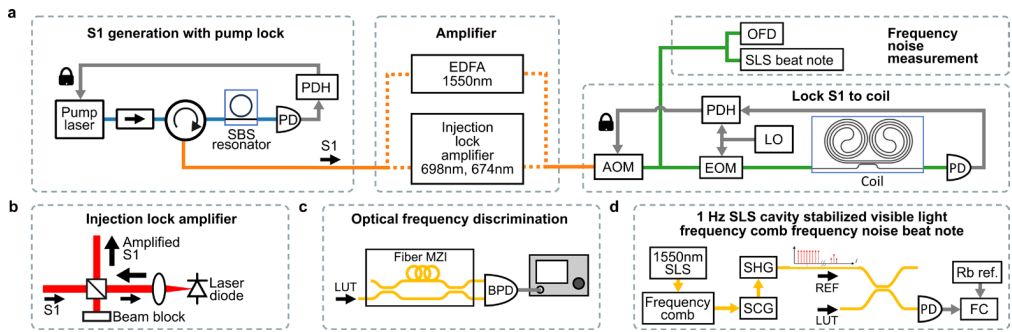

**Fig. 3. Integrated Coil stabilized Brillouin lasers and measurement systems. a** SBS laser locked to coil resonator and frequency noise measurement. **b** Amplifiers. Injection lock to high-power laser diodes for 674 and 698 nm. Commercial Erbium-Doped Fiber Amplifier (EDFA) for 1550 nm. **c** Frequency noise measurement for > 1 kHz with unbalanced fiber MZI as an optical frequency discriminator. LUT: light under test. **d** Frequency noise measurement for 1 Hz to 1 kHz frequency offset using 1Hz visible light optical frequency comb with heterodyne detection beat notes. SLS: stabilized laser system. SCG: Supercontinuum Generation. SHG: second-harmonic generation. FC: frequency counter.

The SBS lasers are designed for mW level threshold operation with integer multiple of the resonator FSR chosen to phase match the Brillouin Stokes shift. The radius of the SBS resonator is 3.73 mm to match the 25.1 GHz Brillouin gain shift at 674 nm[20], 3.82 mm to match the 23.3 GHz Brillouin gain shift at 698 nm, and 11.83 mm to match 10.9 GHz Brillouin gain shift at 1550 nm[26]. The S1 thresholds are measured to be 7 mW, 5 mW, and 15 mW at 674 nm, 698 nm, and 1550 nm[26] respectively (Supplementary Information S2). The threshold varies with the propagation loss, resonator bus coupling, and FSR of the cavity relative to the Stokes frequency shift[37]. Shorter wavelengths have higher propagation loss resulting in lower Q and higher thresholds for VIS operation. The coupling condition of the cavity affects both the threshold and slope efficiency of the Brillouin lasing. Stronger coupling between the bus waveguide and the resonator yields a higher slope efficiency and a higher threshold. Mismatch of integer multiples of the FSR from the peak Brillouin gain shift for a particular wavelength operation will also increase the S1 threshold.

To generate SBS, the pump laser is PDH locked to the SBS cavity, and the first order Stokes (S1) is extracted in the counter propagating direction using a circulator. The SBS laser is operated below the second order Stokes (S2) threshold to avoid cascaded emission of SBS which degrades the S1 fundamental linewidth[37]. The fundamental linewidths are calculated from the measured white frequency noise (typically above 1 MHz offset), and are reduced to 14 Hz for 674 nm, 7 Hz for 698 nm, and 1 Hz for 1550 nm. This represents a fundamental linewidth reduction from the pump to S1 of 3300X, 280X, and 96X respectively.

For the visible wavelengths, a 3-meter length coil resonator design is used that consists of a waveguide bus connected to two inter-connected 1.5 m spirals, resulting in an optical mode volume that is approximately 125 times that of the 674 nm and 698 nm SBS resonators. At 1550 nm, we use a 4-meter-long coil design with a similar layout to the 3 m coils. We use an AOM to frequency shift the SBS laser output for stabilization to the integrated coil resonator cavity (Fig. 3a) with an optical amplifier to compensate for the AOM optical loss. For amplification at 674 nm and 698 nm we use injection locked high-power laser diodes (Ushio HL67001DG, HL70021DG) and at 1550 nm we use a commercial EDFA (Fig. 3a and 3b). In Fig. 3, two main paths are shown, the locking path and measurement path. The locking path modulates the SBS output (S1) using an electro-optic modulator (EOM) for stabilization to the coil resonator. The coil resonator transmission port is then detected for PDH locking. The PDH servo correction is fed back to the AOM to perform frequency correction. The measurement path is used to characterize the frequency noise using an unbalanced fiber MZI and beat note measurement with a ULE stabilized frequency comb and SHG. The unbalanced fiber MZI (Fig. 3c) is used to measure frequency noise in the ~1kHz to 10 MHz range (see Methods). At low-frequency offsets, 1 Hz to 1 kHz, the fiber MZI environmental fluctuations and noise dominate the laser noise. We use a Vescent fiber frequency comb (model FFC-100) stabilized to a 400,000 finesse ULE cavity (Stable Laser Systems model 6020-4), and frequency-double the comb using SHG to provide a 1 Hz beat note reference from 1800 nm down to 600 nm with 100 MHz comb line frequency spacing. The noise is measured by taking the heterodyne beat note signal between the laser under test and the nearest comb line (Fig. 3d) followed by a frequency counter as described in Methods. The frequency noise data from the heterodyne beat note and OFD are stitched together providing laser noise measurements from 1 Hz to 10 MHz. This data is used to extract the FLW and ILW at all wavelengths reported here.

The resulting stabilized laser noise and linewidth performance are summarized in Fig. 4. We measure a 17 Hz FLW and 322 Hz ILW at 674nm (Fig. 4a), a 17 Hz FLW and 630 Hz ILW at 698nm (Fig. 4b), and a 1.0 Hz FLW and 181 Hz ILW at 1550 nm (Fig. 4c). The frequency noise data before the stitch are shown in Supplementary Information S3. These results are for over an octave span with the ILW reduced by 10x-800x compared to the pump laser and the FLW reduced by 96x-2700x. We note that the fundamental linewidth is slightly increased from S1 due to the added noise of the amplifiers. The ADEV is calculated from the frequency noise data (Fig. 4d) with a minimum of $6.5 \times 10^{-13}$ at 1 ms for 674 nm, $6.0 \times 10^{-13}$ at 15 ms for 698 nm, and $2.6 \times 10^{-13}$ at 15 ms for 1550 nm. The simulated TRN-limited integral linewidth of the coil resonators is below 15 Hz which represents the achievable ILW for all lasers given optimal PDH lock performance.

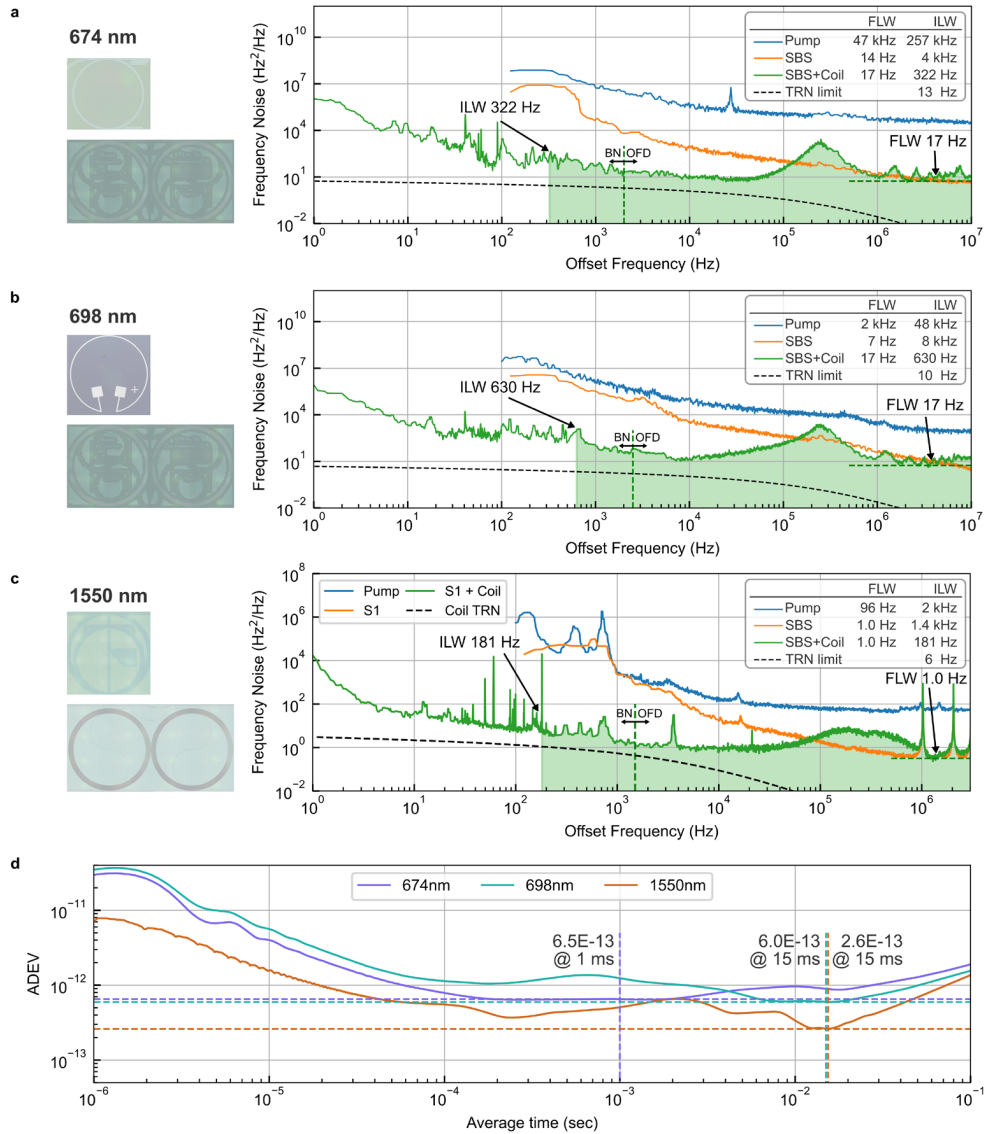

**Fig. 4. Frequency noise of the integrated coil-stabilized Brillouin lasers.** Comparison of the frequency noise of the pump laser, the S1 tone of SBS, and the coil stabilized laser at **a** 674 nm, **b** 698 nm, and **c** 1550 nm. The pump and S1 frequency noise are measured with OFD. The coil-stabilized laser frequency noise is stitched between OFD and heterodyne detection to an ultra-stable reference laser. The stitching points are marked in the plots with vertical dashed lines. The green horizontal dashed lines show the white noise floor determining the fundamental linewidth. The shaded areas show the integral of the frequency noise that determines the integral linewidth of the laser. **d** Allan deviation (ADEV) of the coil stabilized Brillouin laser at 674 nm, 698 nm, and 1550 nm. The dashed lines label the minimum of the ADEV.

## Discussion

We demonstrate silicon nitride integrated coil resonator stabilized Brillouin laser designs that are capable of operating over an octave span from the VIS to SWIR and report record-low linewidths and frequency noise from 1 Hz to 10 MHz with close to TRN limited performance over this wide wavelength range. This dual-stage laser design is demonstrated at key wavelengths in the visible for the strontium atomic species quantum applications and in the SWIR for fiber sensing and ultra-low phase noise microwave and RF generation. With further optimization of the PDH lock[38] and improved acoustic isolation of the integrated laser and coil the linewidths can be reduced to the coil TRN limited 13 Hz, 10 Hz and 6 Hz ILW at 674, 698, and 1550 nm respectively. The fundamental linewidth can be further reduced through improved design of the SBS laser. For example, second-order Stokes (S2) suppression[39] or an increase in the SBS cavity length[40]. Since the Brillouin laser and coil resonator are fabricated on the same silicon nitride platform using the same process flow, the laser and cavity can be integrated on the same chip and combined with other components, for example a 50/50 splitter, S1 optical filter, and pump lasers[21,41].

The integrated SBS laser and coil resonators can be readily implemented at many wavelengths within the $Si_3N_4$ transparency window[31]. Demonstrations at 674 nm and 698 nm specifically target transitions relevant to strontium trapped ions and neutral atoms, making this approach suitable for optical clocks and quantum information processing. The narrow laser linewidth, provides low phase noise at key frequency offsets, which is of significance to the fidelity of qubit operations and stability of atomic clocks[15,34]. The performance of fiber interferometric sensing improves with decreased phase noise[8,9] and the microwave phase noise in optical frequency division approaches can be reduced using low noise stabilized lasers[4,6,7].

Looking ahead, this coil stabilized SBS laser design fabricated in the versatile ultra-low-loss silicon nitride platform provides a CMOS foundry compatible way to integrate a wide range of precision technologies at the chip scale. By combining on-chip SBS and coil resonators with integrated components such as external-cavity tunable lasers (ECTLs)[41–43], piezoelectric (PZT) modulators[44,45], AOMs[46,47], and ion or atom traps[48–50], a complete optical system-on-chip can be realized. These results unlock the potential for compact atomic clocks, scalable quantum computing architectures, high-resolution spectroscopy, low-phase-noise mm-wave generation, and field-deployable sensors.

## Methods

### Fabrication Process

The fabrication starts with 15 μm thermally grown silicon dioxide on top of the silicon substrate. A stoichiometric silicon nitride film is then deposited with low-pressure chemical vapor deposition (LPCVD). The waveguide is patterned by a 248 nm DUV stepper and etched by inductively coupled plasma reactive ion etch with $CF_4/CHF_3/O_2$ gas. The upper cladding is 6 μm silicon dioxide deposited with tetraethylorthosilicate (TEOS)-based plasma enhanced chemical vapor deposition (PECVD) and annealed to mitigate scattering and absorption losses.

### Quality factor measurements and calculation

The power of a frequency-detuning laser is split into two fractions. One fraction is sent through an unbalanced fiber Mach-Zehnder interferometer (MZI) with a calibrated free spectral range (FSR) as a frequency reference. The other fraction is sent through the tested resonator. Both power transmissions of the fiber MZI and the tested resonator are synchronized and recorded with an oscilloscope. The loss and Q of the resonance are calculated by fitting the transmission spectrum to waveguide resonator theory. The measured frequency detuning is calibrated to the FSR of the fiber MZI.

**Optical frequency discrimination and beat note measurements of frequency noise**

Frequency noises are measured either by optical frequency discrimination (OFD) with a fiber MZI or beat note (BN) with a stabilized laser system (SLS) reference. The unbalanced fiber MZI is first calibrated for free-spectral range (FSR) frequency. We send the laser to be tested through the fiber MZI and measure the difference in output optical power with balanced photodetectors. Then, we ramp the phase shift on one arm of the fiber MZI to get peak to peak voltage of the photodetector output. With the calibrated FSR and measured peak-to-peak voltage, the power output change can correspond to an optical frequency shift around the quadrature point. While taking a measurement, we trigger the measurement around the quadrature point and sample with different rates for different offset ranges. In the end, we average the data over several measurements and stitch different sampling rate traces for the full range frequency noise data.

For low-offset frequency noise, which is limited by the fiber MZI vibration and acoustic noises, we measure the beat note between the tested laser and our frequency comb locked to the SLS. The comb operates around the C-band. Therefore, the comb is frequency-doubled to reach the visible wavelength range. We mix the two lasers and measure the beat note signal with a frequency counter. The stability of the SLS allows us to measure the frequency noise in offsets of 1-1000 Hz. Due to the limited optical power of the comb and the nonlinear efficiency of the frequency doubling, at 674 nm and 698 nm, we phase lock the beat note signal to a voltage-controlled oscillator (VCO) to improve the signal-to-noise ratio for the frequency counter.

**Data availability.**

The data and code that support the plots and other findings of this study are available from the corresponding authors upon request.

## Acknowledgments

This material is based upon work supported by DARPA GRYPHON, under Award Number HR0011-22-2-0008, and by the U. S. Army Research Office under contract/grant number W911NF2310179, and by the NSF under Award Number 2016244. The views and conclusions contained in this document are those of the authors and should not be interpreted as representing official policies of DARPA or the U.S. Government. The samples were fabricated in the UCSB Nanofabrication Facility, an open-access laboratory, and Honeywell CMOS foundry.

## Author contributions

M.S., N.C., M.W.H., N.M., K.L., A.S.H., A.I., R.J.N, and D.J.B. prepared the manuscript. N.C. designed the devices at visible wavelengths. M.W.H. and K.L. designed the devices at C-band. M.S., N.C., N.M., C.C and R.J.N. built the system and took the measurements at visible wavelengths. M.W.H. built the system and took the measurements at C-band. M.S., A.I., A.S.H., and K.L. measured the beat note at visible wavelengths. D.J.B. supervised and led the scientific collaboration.

## Conflict of Interest

Dr. Blumenthal's work has work funded by ColdQuant d.b.a. Infleqtion. Dr. Blumenthal has consulted for Infleqtion, received compensation, and owns stock.